\documentclass{aastex}
\usepackage{spr-astr-addons}
\usepackage{url}
\pdfoutput=1

\RequirePackage{color}
\def\imagei{\centerline{\color[gray]{.75}\rule{\hsize}{4pc}}}%
\def\imageii{\centerline{\color[gray]{.75}\rule{4pc}{4pc}}}%

\newcommand{\grado}{\mbox{$^{\circ}$}} 

\newcommand{\teff}{\mbox{$T_{\rm eff}$}}

\newcommand{\as}{\mbox{$^{\prime\prime}$~}}

\newcommand{\rv}{$R_V$~}

\newcommand{\beqa}{\begin{eqnarray}}
\newcommand{\eeqa}{\end{eqnarray}}

\newcommand{\ebv}{\mbox{$E_{B\!-\!V}$}}

\newcommand{\Msun}{\mbox{$M_{\odot}$}}

\begin{document}

\title{The Galaxy Evolution Explorer (GALEX).\\ Its legacy of UV surveys, and science highlights}
\shorttitle{Review of GALEX results}
\shortauthors{Luciana Bianchi}

\author{Luciana Bianchi}
\affil{Dept. of Physics \& Astronomy, The Johns Hopkins University, 3400 N. Charles St.,  Baltimore, MD 21218, USA\\
http://dolomiti.pha.jhu.edu}


\begin{abstract}

The Galaxy Evolution Explorer (GALEX) imaged the sky in the Ultraviolet (UV) for almost a decade, 
 delivering the first sky surveys at these wavelengths. Its database contains
 far-UV (FUV, $\lambda$$_{eff}$ $\sim$ 1528\AA) and near-UV (NUV, $\lambda$$_{eff}$ $\sim$ 2310\AA) images of most of the sky, 
including deep UV-mapping of extended galaxies,  over 200~million source measurements,
and more than 100,000 low-resolution UV spectra.   
The GALEX archive will remain a long-lasting resource for statistical
 studies of hot stellar objects, QSOs, star-forming galaxies, nebulae and the interstellar medium.
It provides an unprecedented 
road-map for planning future UV instrumentation
and follow-up observing  programs in the UV and  at other wavelengths.

We review the characteristics of the GALEX data, and describe final 
catalogs and available tools, 
that facilitate future exploitation of this database. 
We also recall highlights from the science results uniquely enabled by GALEX data so far. 
  

\end{abstract}



\section{Introduction.  GALEX instrument and data}
\label{s_galex}

The Galaxy Evolution Explorer (GALEX),  a NASA {\it Small Explorer} Class mission, was
launched on April 28, 2003 to perform the first sky-wide Ultraviolet surveys, with both direct imaging and 
grism  in two broad bands,  
 FUV ($\lambda$$_{eff}$ $\sim$ 1528\AA, 1344-1786\AA) and  
 NUV ($\lambda$$_{eff}$ $\sim$ 2310\AA,  1771-2831\AA).
It was operated with NASA support until 2012, then for a short time 
with support from private funding from institutions; it was decommissioned on June 28, 2013.
\footnote{GALEX was developed by NASA with contributions from the Centre National d'Etudes Spatiales 
of France and the Korean Ministry of Science and Technology.}

\begin{figure*}
\label{f_figure1}
\vskip -.72cm
\centerline{
\includegraphics[width=6.5cm]{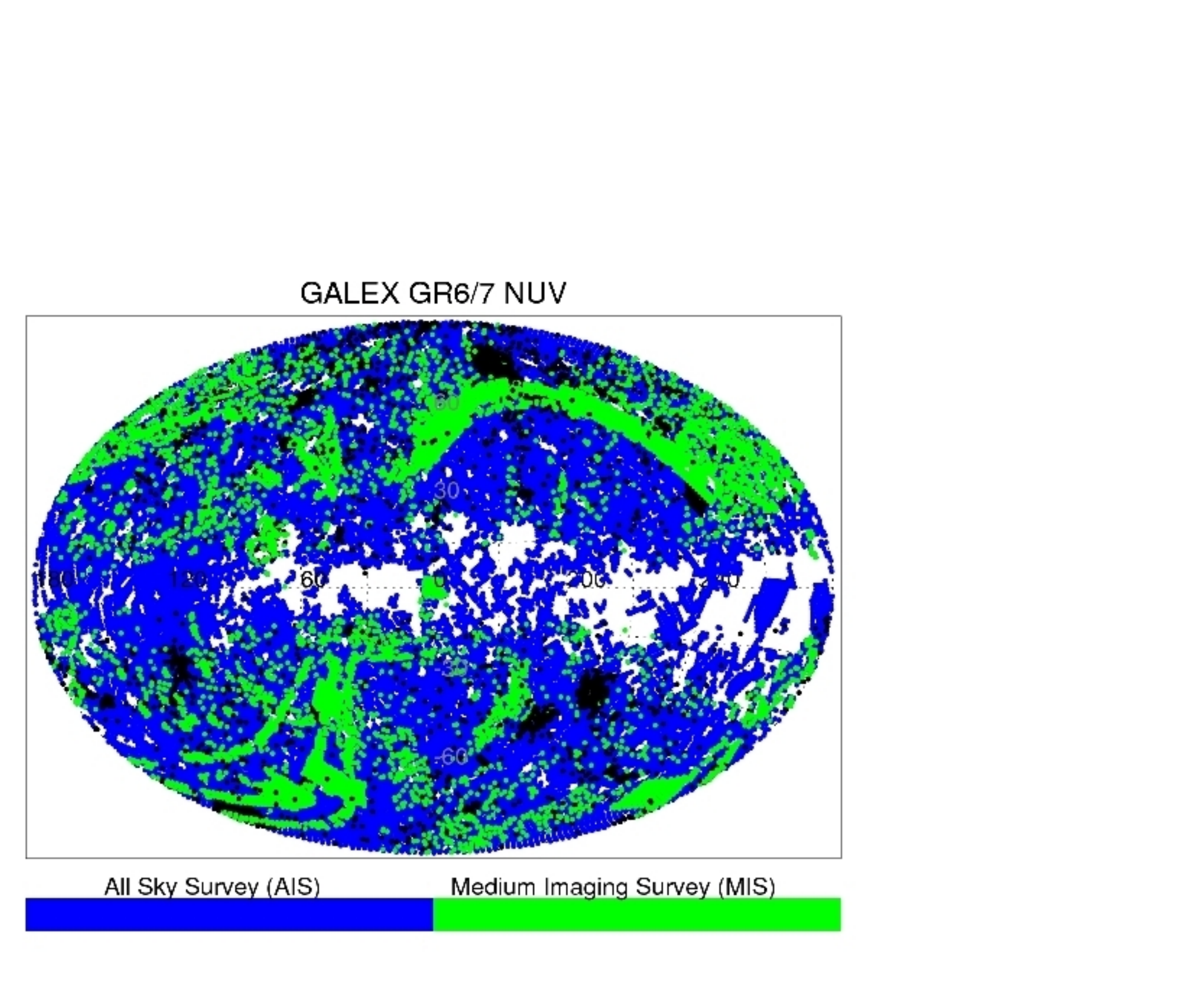}
\includegraphics[width=6.5cm]{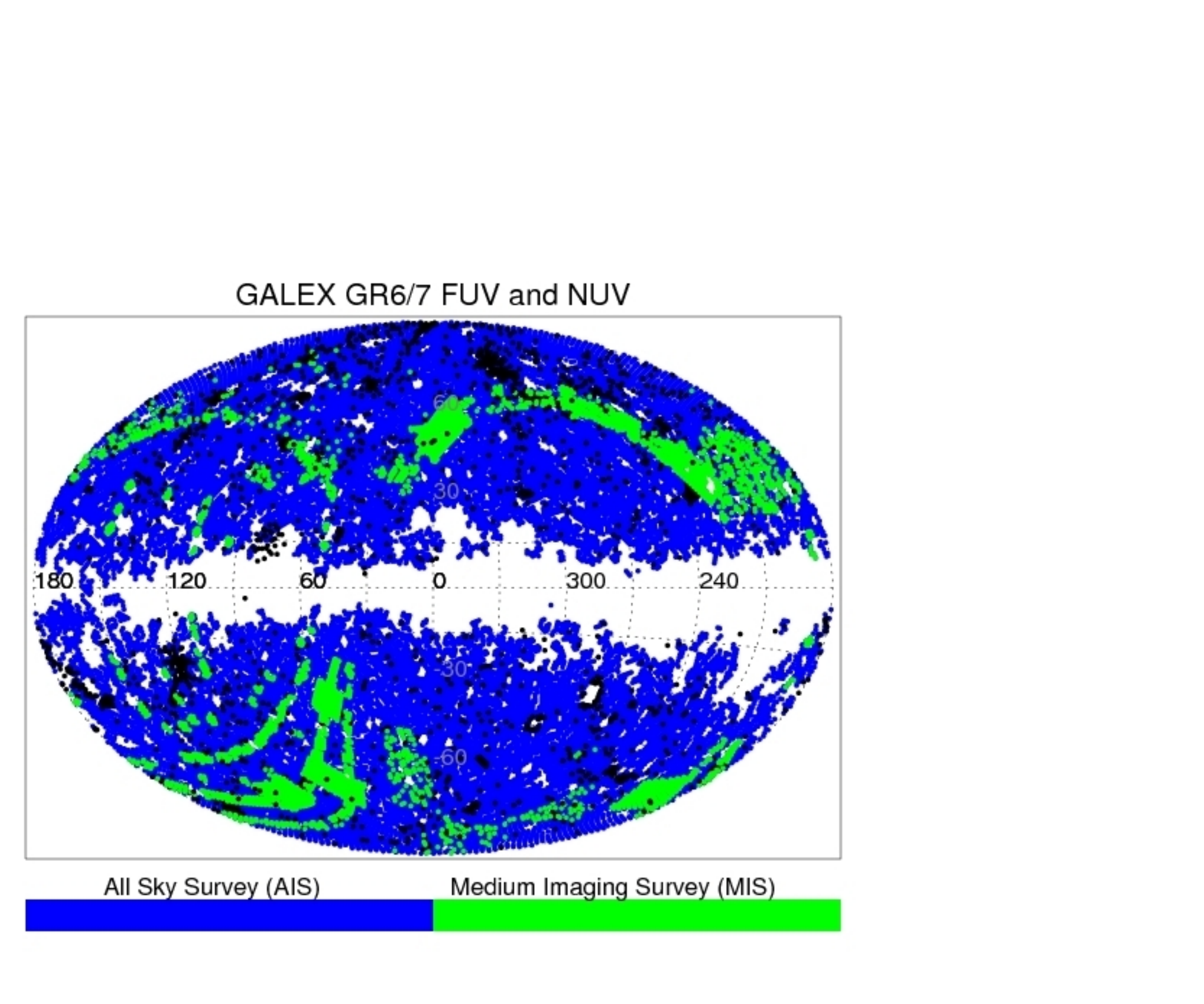}}
\vskip -.62cm 
\caption{Sky coverage, in Galactic coordinates, of the GALEX imaging. 
The surveys with the largest  area coverage  are 
 AIS (blue) and  MIS (green). Observations from other surveys are shown  in black (figure
adapted from Bianchi et al. 2014a).
Data from the privately-funded observations at the end of the mission  are not shown. 
Left: fields observed with at least the NUV detector on; right:
fields observed with both FUV and NUV detectors on. The latter constitute the BCScat's.}
\end{figure*}

GALEX's instrument  consisted of a  Ritchey-Chr\'etien$-$type telescope, with a 50~cm primary mirror and a  focal length of
299.8cm.  Through a dichroic beam splitter, light was fed  to the FUV and NUV detectors simultaneusly. 
The FUV detector stopped working in May 2009; subsequent  GALEX observations have only
 NUV imaging (Figure \ref{f_figure1}). 

The GALEX field of view 
is $\approx$1.2$^{\circ}$  diameter (1.28/1.24$^{\circ}$, FUV/NUV), and the spatial resolution is 
 $\approx$ 4.2/5.3\as (Morrissey et al. 2007). 
 For each observation, the photon list recorded by the two photon-counting micro-channel plate 
detectors
is used  to reconstruct an  FUV and an NUV image, 
 sampled with  virtual pixels of 1.5\as.  
From the reconstructed image, the  pipeline then derives a 
sky background image, and performs source photometry.
Sources detected in FUV and NUV images of the same observation
 are matched by the pipeline with a 3\as radius, to produce a merged list for 
each observation (Morrissey et al. 2007). 

 To  reduce localized 
response variations, in order to 
maximize photometric accuracy, each observation was carried out
 with a 1$^{\prime}$ spiral dithering pattern.
The surveys were accumulated by painting the sky with contiguous {\it tiles}, with series of such observations.
A trailing mode was instead used for the latest,  privately-funded observations,
to cover some bright areas near the MW plane.  These latest data currently are not in the public archive.

At the end of the GALEX mission,  the AIS survey was extended  towards 
the Galactic plane, largely inaccessible during the prime mission phase because of the 
many bright stars that violated high-countrate safety limits. A survey of 
the Magellanic Clouds (MC), also previously unfeasible due to brightness limits, 
was completed at the end,   relaxing the initial count-rate saftey threshold.  Because of the 
FUV detector's failure, these extensions include only
NUV measurements (Figure \ref{f_figure1}). 

\begin{figure*}
\label{f_crowd}
\vskip -.5cm
\centerline{
\includegraphics[width=15cm]{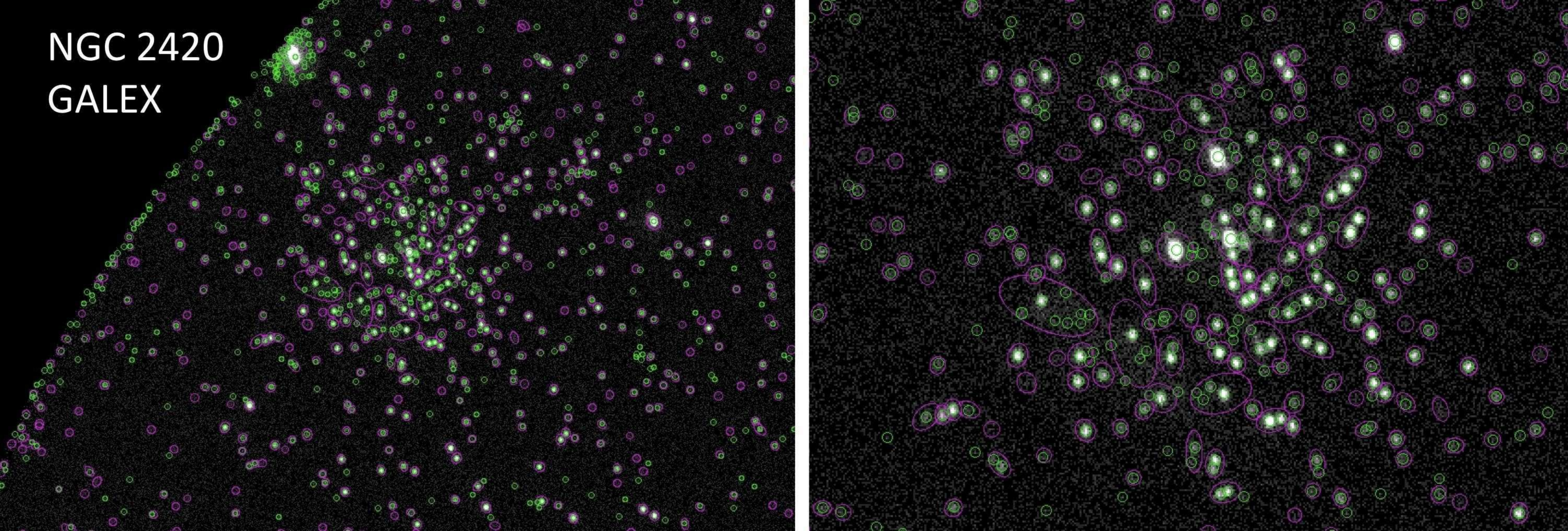}}
\caption{Portion of a GALEX field with the stellar cluster NGC2420 (de Martino et al. 2008). 
Green circles mark sources detected with our photometry, the purple contours mark sources
as defined by the GALEX pipeline.  The right-side image is an enlargement of the crowded region.
The left image also includes a section of 
the field's outer edge,  showing how numerous rim artefacts intrude photometric source
detections.  The rim is excluded from the BCS catalogs. }
\end{figure*}

\begin{figure*}
\label{f_6822}
\centerline{
\includegraphics[width=17cm]{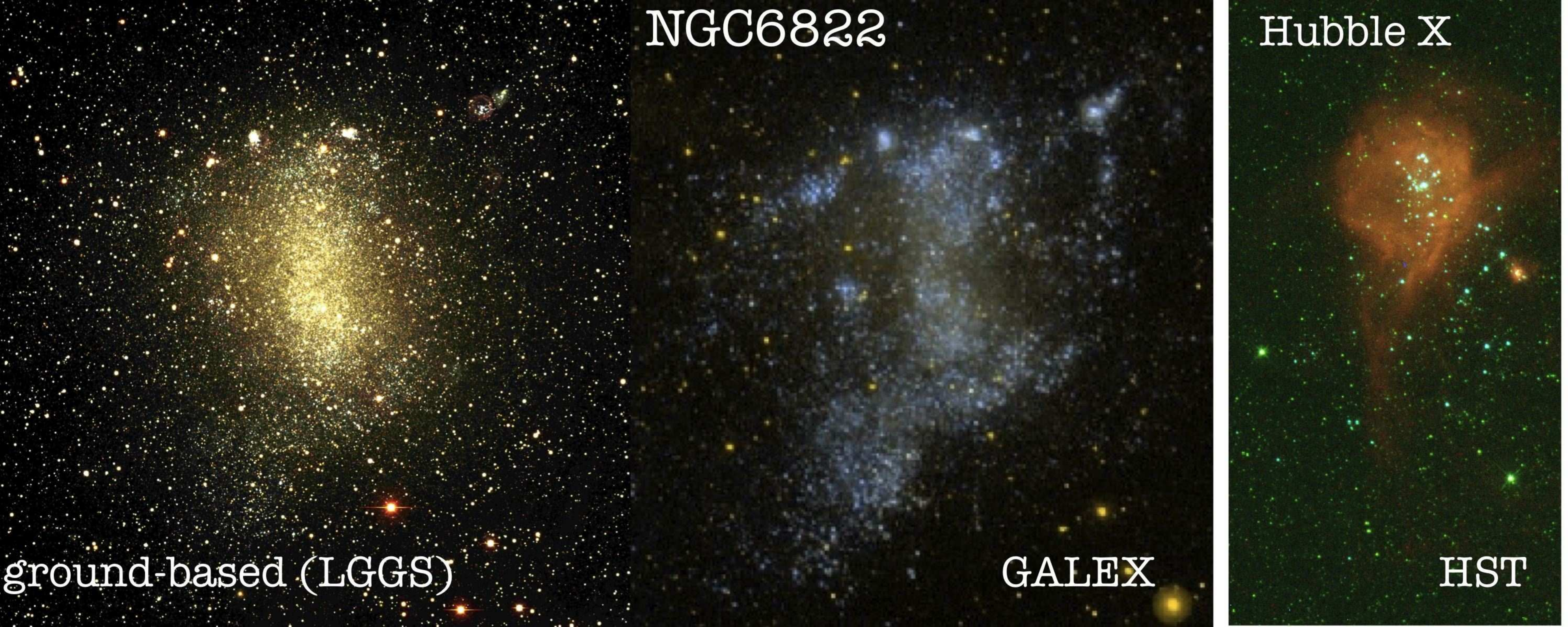}}
\caption{GALEX images (FUV: blue, NUV: yellow) and optical color-composite images of  
  NGC6822 
in the Local Group 
(Efremova et al. 2011, Bianchi et al. 2011c, 2012) and  HST view
of one of its most prominent HII regions, Hubble~X (Hubble data from Bianchi et al. 2001). Hubble~X 
is one of the two bright knots in the upper part of the galaxy in GALEX and ground-based images; its core is resolved into an
association of young stars with HST  (0.1\as resolution, or $\approx$0.2pc at a distance of 460~kpc).}
\end{figure*}

Much of this short review concerns 
 GALEX's final data products (from the point of view of science applications;
technical documentation is available elsewhere),
This choice, at the price of 
 confining science results to a few highlights (Section \ref{s_science}), 
 responds to various inquiries and requests, timely as the almost entire (and final) database is becoming available, 
and new tools, which will support new investigations.

\section{GALEX surveys}
\label{s_surveys}

GALEX has performed sky surveys with different depth and coverage (Morrissey et al. 2007, Bianchi 2009).
The two detectors, FUV and NUV,  observed simultaneously 
as long as the FUV detector
was operational; note however that there are occasional observations in which one of the two detectors
was off (mostly FUV) due to brief shut-downs, even in the early part of the mission, and in some
observations the FUV and NUV exposure times differ (see Bianchi et al. 2014a, in 
particular their Table 1 and Fig. 2). 

 The surveys with the largest area coverage are the All-Sky Imaging survey (AIS) 
and the Medium-depth Imaging Survey (MIS). Exposure times slightly vary within each survey,
around the respective 
 nominal exposures of 100~sec for AIS (corresponding to a depth of FUV$\sim$20/NUV$\sim$21~ABmag)
and 1500~sec for MIS (corresponding to a depth of $\sim$22.7~ABmag in both FUV and NUV). 
The Deep Imaging Survey (DIS) accumulated exposures of the order of several tens of thousand of seconds in selected fields
(for a 30,000~sec exposure, the depth reached is $\sim$24.8/24.4~ABmag in FUV/NUV).  
The  ``Nearby Galaxies Survey'' (Bianchi et al. 2003, Gil de Paz et al. 2007) covered initially 436 fields at 
MIS depth, but hundreds of additional nearby galaxies 
 were mapped by GALEX, as part of MIS or other surveys.
Other observations were obtained for guest investigator (GI) programs, and for other targeted regions  such as, 
for example, the Kepler field. 

 The  GALEX database at the end of the mission  (2013, data release GR7)  contained 214,449,551 source measurements, most of which 
(210,691,504)  from observations with both 
detectors on. However, a small number of additional datasets obtained throughout the mission,
which were not previously included in the archive, were now reprocessed as part of the MAST effort to 
finalize the database, therefore the final archive will contain a few more entries. 
Figure \ref{f_figure1} shows the sky coverage of all GALEX observations 
performed in both FUV and NUV (right), and in NUV regardless of FUV-detector status (left). 
The map does not include the last NUV trailed observations (CLAUSE dataset).

\section{GALEX database and science catalogs}
\label{s_CBS}

GALEX data can be obtained from the NASA MAST archive at http://galex.stsci.edu. They include
images (direct imaging or grism) and associated photometry from the pipeline, or extracted spectra. 
  High-level science catalogs (HLSP) are also available, 
such as those  described below, which greatly facilitate some investigations. 

Several objects have multiple measurements in the database, due to repeated  observations of the same
field, or overlap between fields. For  studies involving UV-source counts, one needs to 
eliminate repeats, as well as artefacts.  Therefore, we have  constructed catalogs of 
{\it unique} UV sources, eliminating duplicate measurements of the same object.
Separate catalogs were constructed for AIS and MIS, because of the $\sim$2-3 mag 
difference in depth. 
The most recent version of such catalogs is published by Bianchi et al. (2014a, hereafter ``BCS'', see Appendix A); BCS 
also presented 
sky maps showing density of UV sources with various cuts.  
A previous version, based on data release 
GR5, was published by Bianchi et al. (2011a, b), who extensively discuss
 criteria for constructing  GALEX source catalogs and matched catalogs between GALEX 
and 
SDSS, GSC2, and more. Earlier work on source classification was
presented by Bianchi et al. (2007, 2005), Bianchi (2009). 

In the GALEX database, the  non-detection value for the FUV magnitude, FUV =-999,
 means  either that the detector was on but the source was too faint to be measured in FUV, 
of the FUV detector was off.  In order to examine and classify sources by colors, and relative 
fraction of sources with different colors,  Bianchi et al. (2014a, 2011a) restricted the catalogs to 
those observations in which both detectors were exposed. 
In addition, these catalogs were conservatively restricted to measurements within the central 1\grado~ diameter of the field of view,  to exclude 
the  outer rim, where distortions 
prevent position and photometry of sources to be derived accurately, and counts from rim's spikes cause
numerous artefacts to intrude the source list.
 For more details we refer to the original BCS  publication. 
 
The BCS catalogs  include  28,707 AIS fields with both FUV and NUV exposed,
covering a unique area of 22,080 square degrees when restricted to the central 1\grado~ of each field,
and 3,008 MIS fields 
(2,251  square degrees). 
An updated version (MIS) will be posted on the same site, including the newly recovered data
which will pass final quality test. 
The area coverage at MIS depth was significantly increased after the FUV detector failed, 
therefore additional data with NUV-only measurements exist (Figure \ref{f_figure1}). 
The BCS catalogs (``BCScat'' in MAST casjobs), for AIS and MIS, include {\bf all GALEX observations with both FUV and
NUV detectors exposed,} up to the latest, and final, release (GR7). They contain $\approx$71 and
$\approx$16.6~Million sources respectively.

\begin{figure}
\label{f_uvsky}
\centerline{
\includegraphics[width=7.5cm]{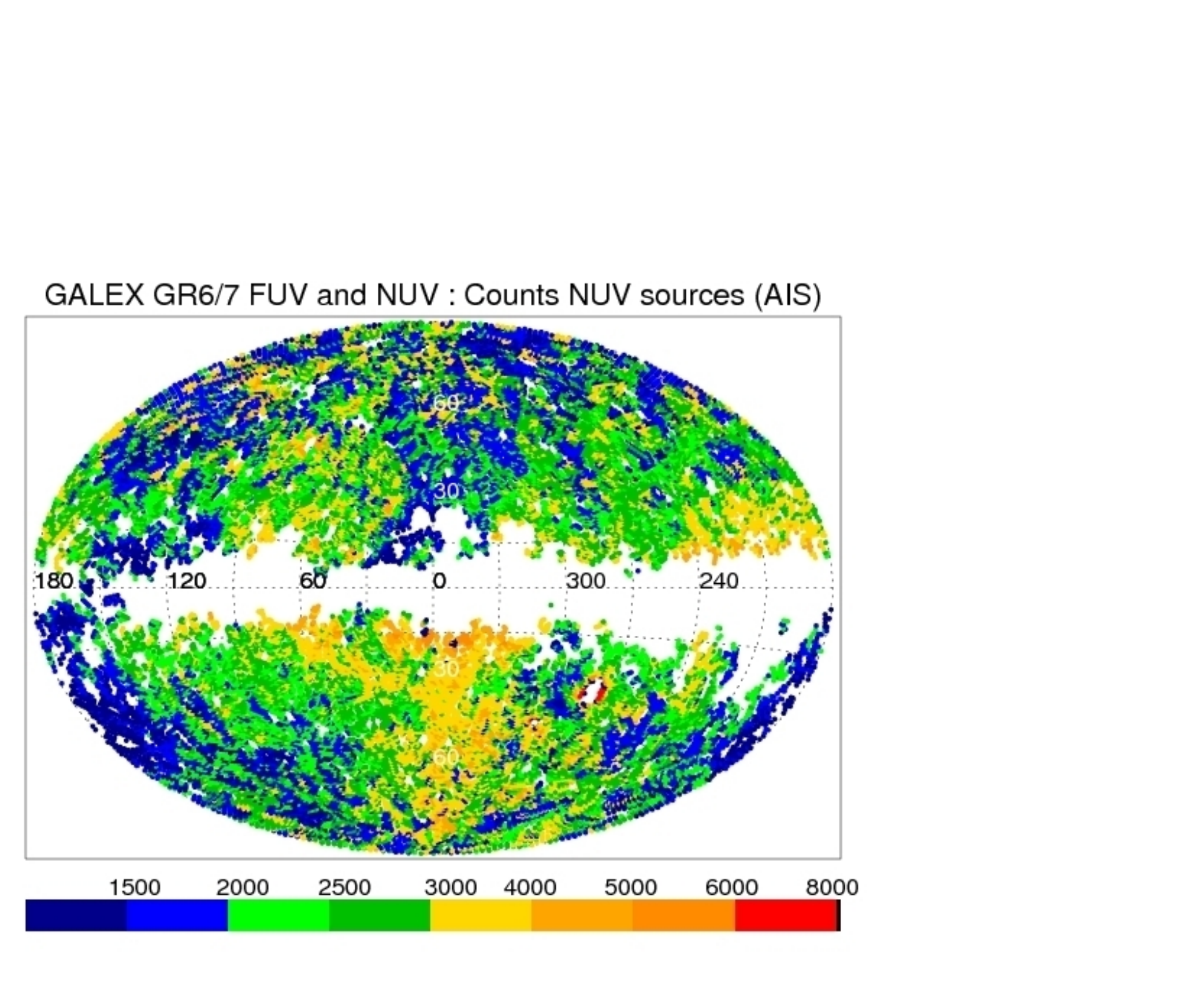}}
\centerline{
\includegraphics[width=7.5cm]{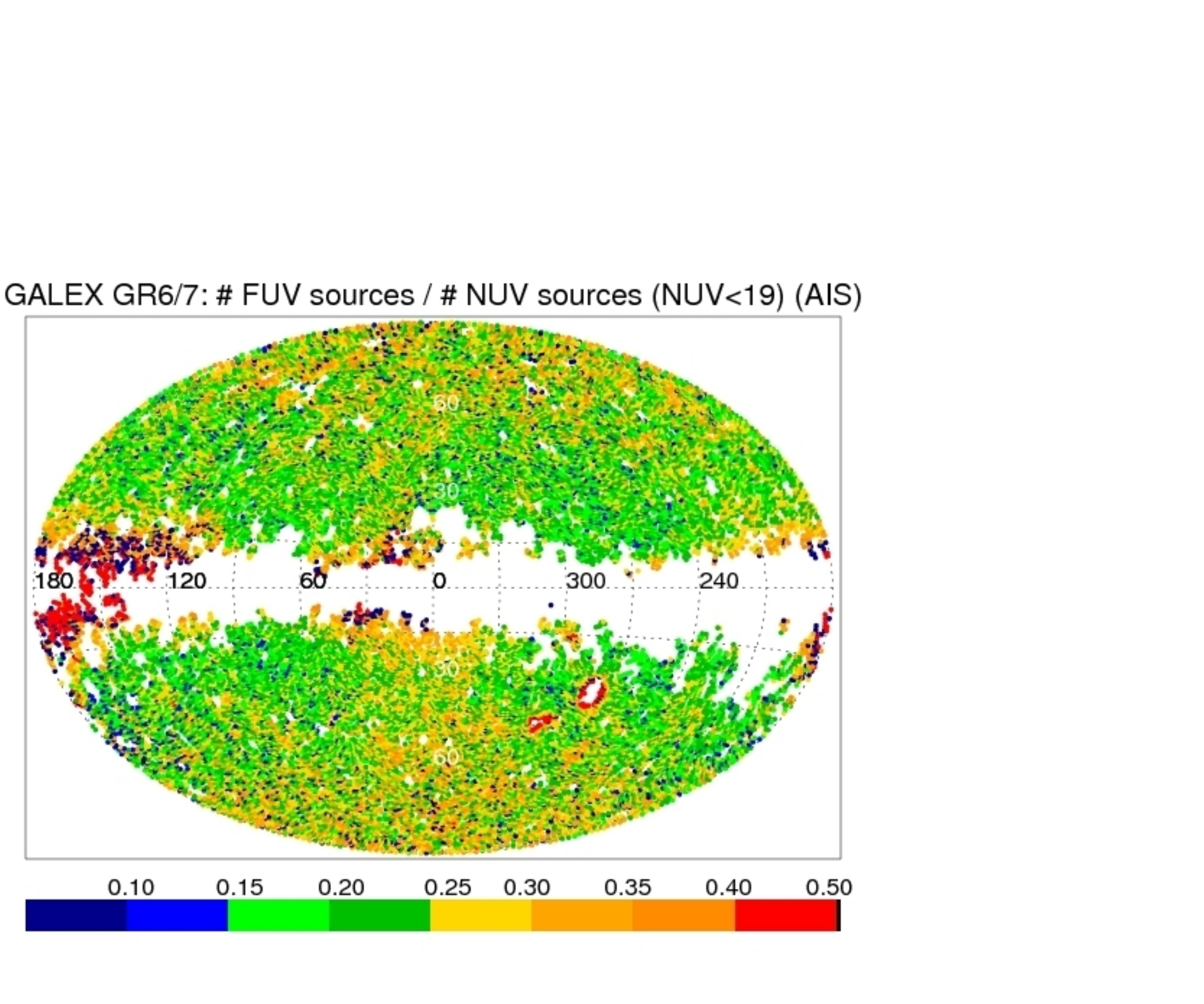}} 
\centerline{
\includegraphics[width=7.5cm]{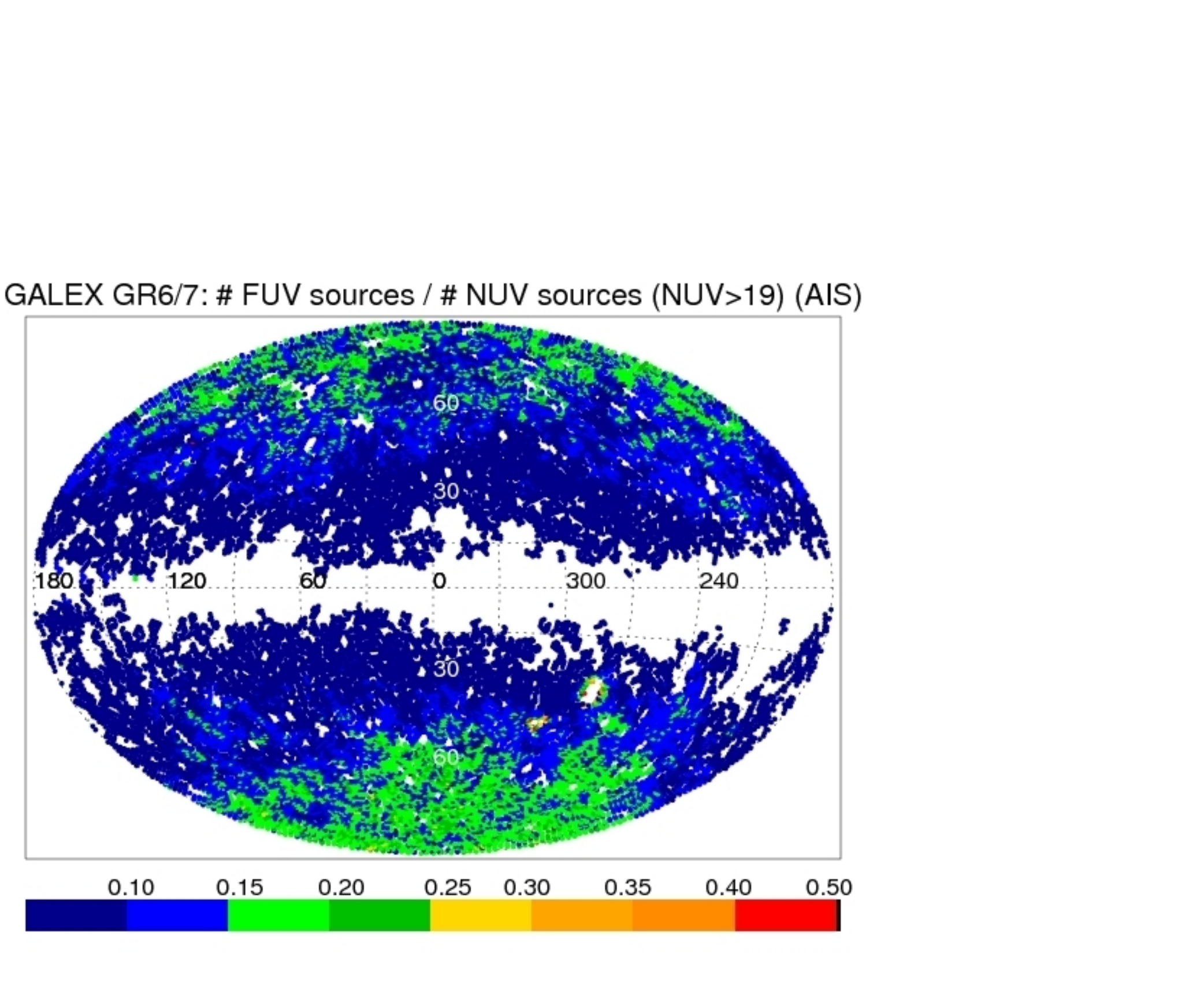}} 
\vskip -.65cm 
\caption{Counts of NUV sources per square degree (top), from the AIS: part of the structure is due
to inhomogeneous exposure times, part is real.  
Fraction of FUV over NUV detections, in the AIS catalogs, for sources brighter (middle)
and fainter (bottom) than NUV=19~ABmag. 
The latter are dominated by extragalactic sources, the brighter sources contain a high
fraction of Milky Way hot stars (see also Bianchi et al. 2011a,b).
Maps in Galactic coordinates,  adapted from Bianchi et al. 2014a)}
\end{figure}

\section{Using GALEX Data: Notes and Caveats}

\subsection{Bright sources} 
 High cout-rates from UV-bright  sources  cause non-linearity in the response, or saturation, mostly 
due to the detector's  dead-time correction. \cite{Morrissey07}  showed that non-linearity,  
 at a 10\% rolloff, sets in  at 109~counts~s$^{-1}$ for FUV and 311~counts~s$^{-1}$ for NUV.
These countrates correspond to FUV=13.73~ABmag ($\sim$1.53~10$^{-13}$ erg~s$^{-1}$~cm$^{-2}$~\AA$^{-1}$)
and NUV=13.85~ABmag ($\sim$6.41~10$^{-14}$ erg~s$^{-1}$~cm$^{-2}$~\AA$^{-1}$). A correction for non-linearity 
is applicable over a limited range, beyond which the measured 
countrates saturate and the true source flux is no longer recoverable (see their Figure 8). 
The bright-object limit during most of the mission was 30,000~counts~s$^{-1}$ per source,
corresponding to $\sim$9$^{th}$ABmag for NUV ($\sim$7~10$^{-12}$ erg~s$^{-1}$~cm$^{-2}$~\AA$^{-1}$)
and 5,000~counts~s$^{-1}$ per source in FUV ($\sim$ 9.6~ABmag, 
$\sim$6~10$^{-12}$ erg~s$^{-1}$~cm$^{-2}$~\AA$^{-1}$). 
Such limits were relaxed at the end of the mission. 

The calibration of GALEX fluxes is tied to the UV standards used for HST \citep{rcb2001wd}.
However, all but one of the white dwarf (WD) stardard stars have GALEX count-rates in the non-linear regime. 
 \cite{camarotaholberg14} derived an empirical correction to the GALEX magnitudes in the non-linear range, 
using a well studied sample of WDs with UV spectra and  models.  Their correction is valid 
in the bright-flux regime as specified in their work, but would diverge if extrapolated to fainter fluxes. 
Possible further refinements 
 \citep{rcb2008wd}
have not yet been explored to our knowledge. 

\subsection{Crowded fields} 
\label{s_crowd}

Source detection and photometry performed by 
the GALEX pipeline become unreliable where sources are too crowded relatively to the instrument's resolution. 
Examples include stellar clusters in the Milky Way (Figure 2), fields in or near the Magellanic Clouds (Section \ref{s_mc}), 
and nearby extended galaxies. 
 We note that, in some crowded fields, even sources with separation 
comparable or larger than the image resolution
are sometimes not resolved; see Figure 2 
as an example, or Figure 3 of Simons et al. (2014) for a 
Magellanic Cloud field. In some cases,  the local background may compound the crowdiness
 around clustered sources.  The pipeline, designed for the general purpose of detecting both point-like
and  extended sources (such as galaxies, with 
an elliptical shape), may interpret two or more nearby point sources 
 as one extended source; this seems to occur in crowded regions, as Figure 2 
shows. 

In extended galaxies, because  UV fluxes are sensitive to the youngest, hottest stars, 
which are typically arranged in compact star-forming complexes (see Fig. 3), 
  UV-emission peaks 
are identified by the pipeline as individual sources. 
Often, more complex measurements are needed in extended
galaxies, with special care to background subtraction (e.g. Kang et al. 2009, Efremova et al. 2011, Bianchi et al. 2011b).

 For consistency, all  measurements in the master database with FUV and NUV data
were used to
produce the GALEX BCS catalogs. Large galaxies, stellar clusters, and MC fields 
 were not excluded, to avoid introducing gaps in the catalogs, because what areas must be excluded depends
 on the specific science application. It is the choice (and
responsibility) of the user to exclude known crowded regions when using large datasets, or  check 
the photometry if such regions cannot be excluded, and use 
specific custom-vetted photometry catalogs for these particular areas when possible.

\subsection{Calculation of survey area coverage}
\label{s_area}

Overlaps or gaps may exist between contiguous fields. When unique-source catalogs are 
used, if field overlaps occur in the subset of choice,  
the exact area coverage must be calculated.  Different tools are or will soon be
available, using heal-pixels or tesserae, in the catalogs' web sites (http://dolomiti.pha.jhu.edu/uvsky for {\it BCScat}).

\section{Variable sources} 
\label{s_var}

Serendipitous variability searches in the GALEX database include essentially two possibilities. On one hand,
one can compare photometry of objects  having
repeated observations,  either in consecutive orbits or spaced in time during GALEX's several 
years of operation. The  time intervals sampled are the gaps between different observations,
and each measurement is integrated over the entire exposure of its observation. 
This random cadence is by necessity not optimized for detecting specific periodicities or time-scales,
but its serendipitous nature
 enables a vaster exploration than what 
a single dedicated program can afford. 
It was exploited by a number of works, 
e.g. Welsh et al. (2006, 2007, 2011), Wheatley et al. (2012), Gezari et al. (2013).  
A catalog of GALEX sources showing conspicuous variations ($\Delta$NUV$>$0.6mag,
$\Delta$FUV$>$0.4mag) has been compiled by Conti et al.(2014). Over 400,000 sources 
with NUV$\leq$21~ABmag were found to
satisfy these criteria, $>$7,000 of which have over 30 measurements; they include RR~Lyrae,
eclipsing binaries, flare stars, QSOs, and more.  

In the near future, it will also be possible to investigate variability on time scales shorter
than the integration time of each observation. 
By the time this paper will go to press, a new tool will be available, 'gPhoton' (on the MAST web site).
 It allows users to analyze the entire photon list from an observation, and to extract
 photometry with user-defined short integration times, within 
 the exposure. Such tool opens new possibilities for studies of short time variations of bright sources, such as the flare from an M-type dwarf discovered by Robinson et al. (2005), that displayed  a
1000-fold increase in FUV flux  within 200~seconds.

\section{GALEX survey of the Magellanic Clouds}
\label{s_mc}

GALEX has 
surveyed  the Magellanic Clouds (MC) and
their surrounding areas including the Magellanic Bridge (MB). The central regions are 
only mapped in NUV; because they are very UV-bright, they were avoided during the 
main mission for risk of detector's damage; later the FUV detector had failed.   The 
coverage is shown in Fig. 5 
(insert): it shows a few more fields than Simons et al. (2014)
as we are trying to recover data that initially did not pass the quality check
and are not yet ingested in the archive. 
The 5$\sigma$ depth of the NUV imaging varies between
20.8 and 22.7~ABmag.  
GALEX imaging provided the first sensitive view of the entire content of hot stars in the
Magellanic System, revealing 
young populations even in sites with extremely low star-formation rate
surface density, such as the MB.
As discussed in Section \ref{s_crowd}, source  crowding limits the quality of source detection and photometry 
from standard pipeline
processing. Therefore, custom PSF-fitting photometry of the GALEX data in the MC survey region  was performed.

There are 
$\sim$400 fields within  $<$10\grado from the SMC  and 
$\gtrsim$1250 (depending on quality cut) 
fields within  $<$15\grado~
from the LMC center,  of which $\gtrsim$865 with AIS-depth exposures, and the rest with longer exposures.
The latter set (384 non-AIS fields 
in the LMC, median exposure of 730~seconds), was  presented 
by Simons et al. (2014). A total of 17~million source detections was reduced to about 11~millions
by excluding sources farther than 0.55\grado~ from the field center, and sources with roundness/sharpness 
deviating by $>$2$\sigma$ from the median PSF. These measurements were merged into a catalog of about 6~million 
unique sources. This subset from the deeper exposures (albeit mostly 
in NUV only) encompasses the classical optical  extent of the LMC. 
The fields trace from  the lowest hot-star density in the periphery to the most crowded inner sites,
from $\sim$430 to $\sim$200,000 stars/kpc$^2$ brighter than NUV=19~ABmag (10$\times$ more if no magnitude cut is applied).

The 
final catalog  (Thilker et al. 2014a,b)
will be posted  incrementally 
on http://dolomiti.pha.jhu.edu/uvsky, together with science analysis results.
 Figure 5 
 shows also an example of  SED model analysis 
 of GALEX sources matched to existing optical data, to
characterize hot stars and dust extinction, and to search for
 evolved stars. Deeper optical observations are planned, to better complement the GALEX catalogs. 

~

\begin{figure}[!ht]
\label{f_mc}
\centerline{
\includegraphics[width=7.3cm]{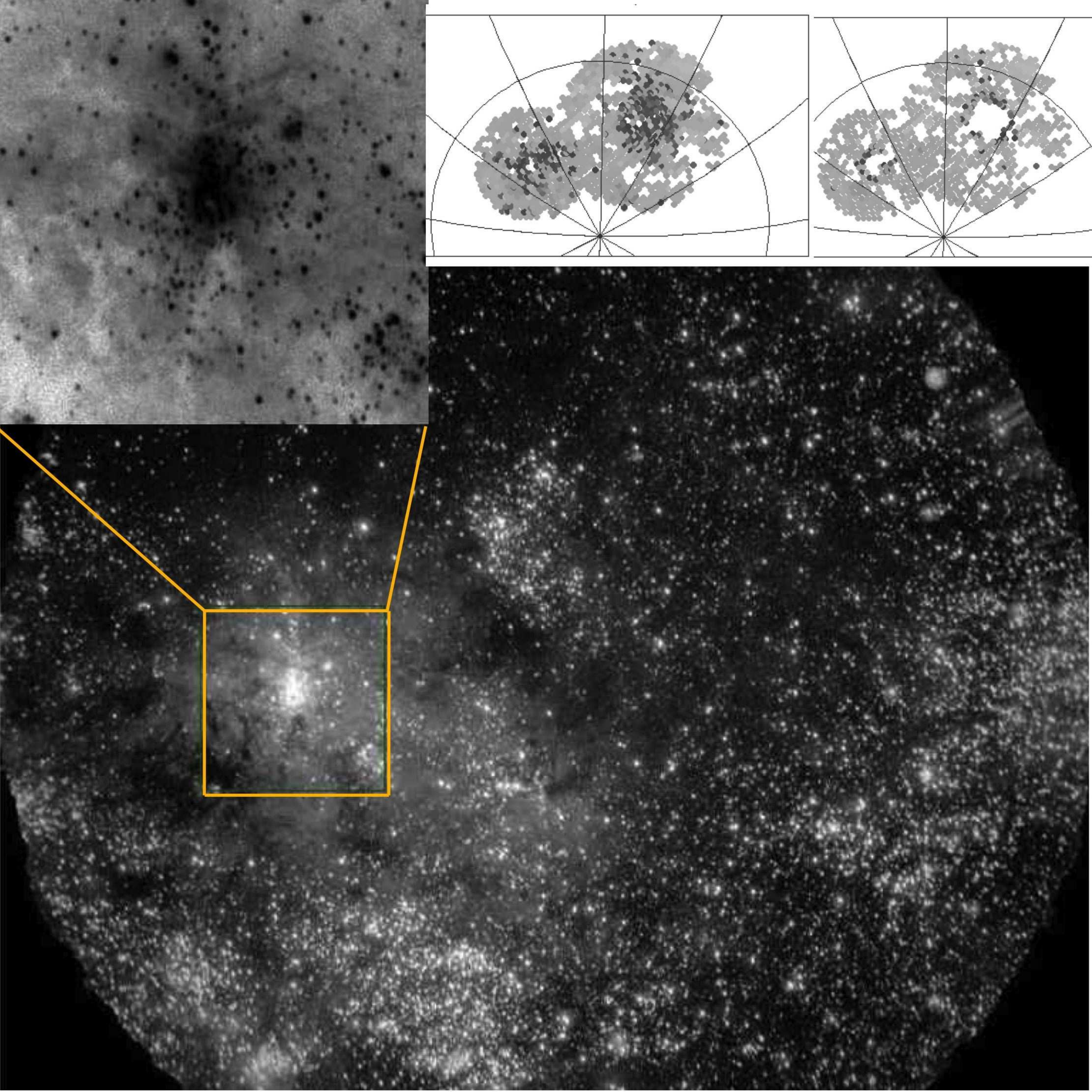}} 
\centerline{
\includegraphics[width=7.3cm]{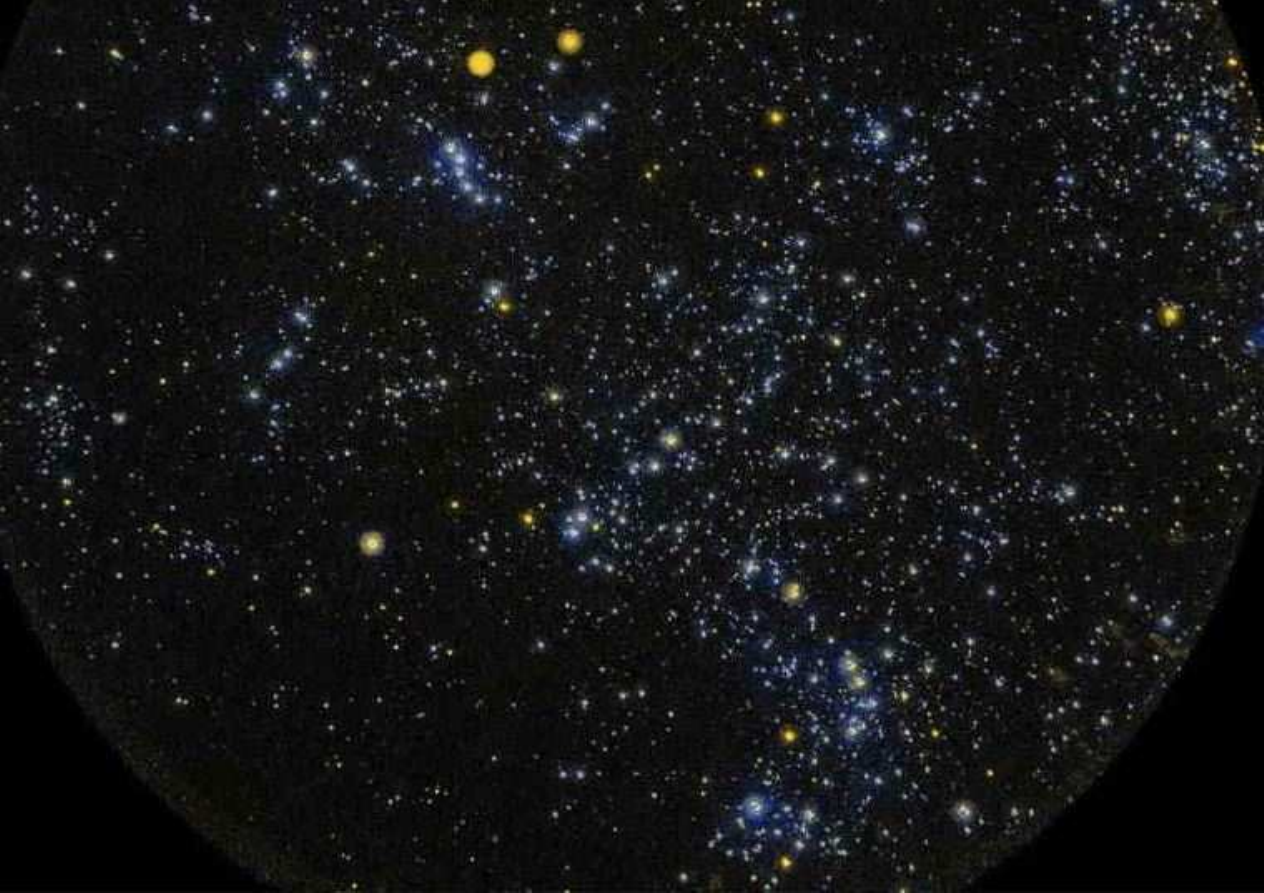}} 
\vskip -2.5cm
\includegraphics[height=2.5cm]{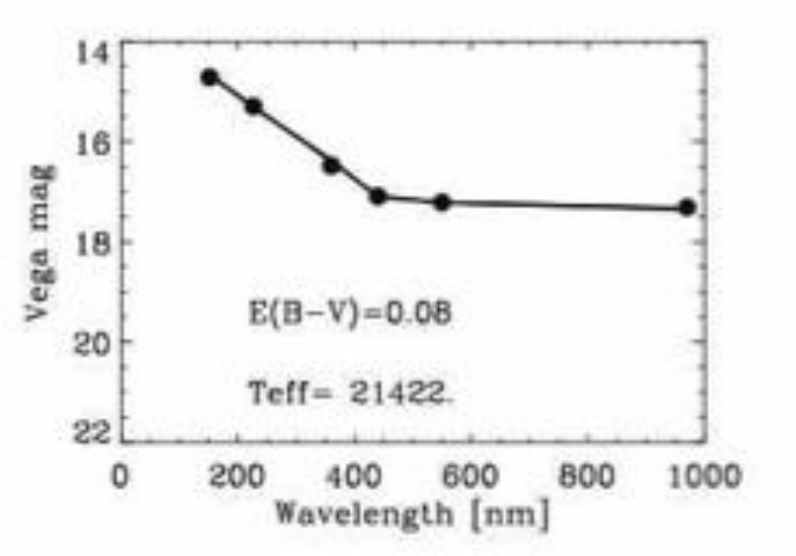}
\caption{The top-right insert shows the location of GALEX MC fields with NUV (left) and FUV (right) observations;
light-to-dark gray from short (AIS-depth) to deeper exposures. The large NUV image is a region (about 1kpc) including
30~Dor, the latter is shown enlarged in the upper-left panel. In addition to  numerous UV-emitting stars,
dust lanes and diffuse emission are also conspicuous. The bottom image is a FUV(blue)+NUV(yellow) composite of a
region in the Magellanic Bridge, where GALEX reveals presence of hot massive stars in areas with gas
density down to N(H)$\sim$10$^{20}$cm$^{-2}$.   The bottom insert is an example of SED-fitting of GALEX+optical photometry. 
}
\end{figure}

\section{Highlights of GALEX science results}
\label{s_science}

\begin{figure*}
\label{f_galaxies}
\vskip -2cm
\centerline{
\includegraphics[width=12cm]{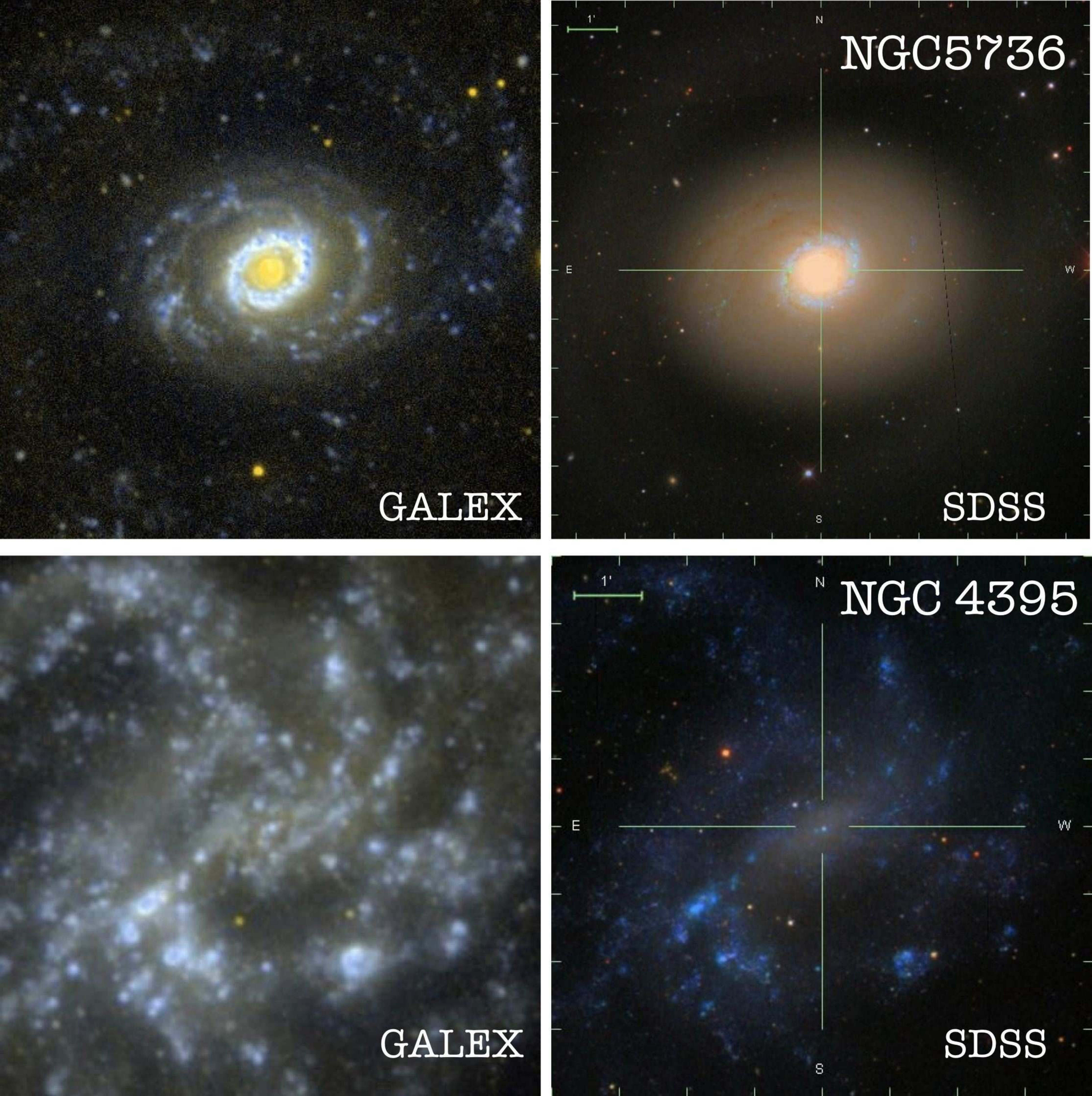}} 
\centerline{
\includegraphics[width=12cm]{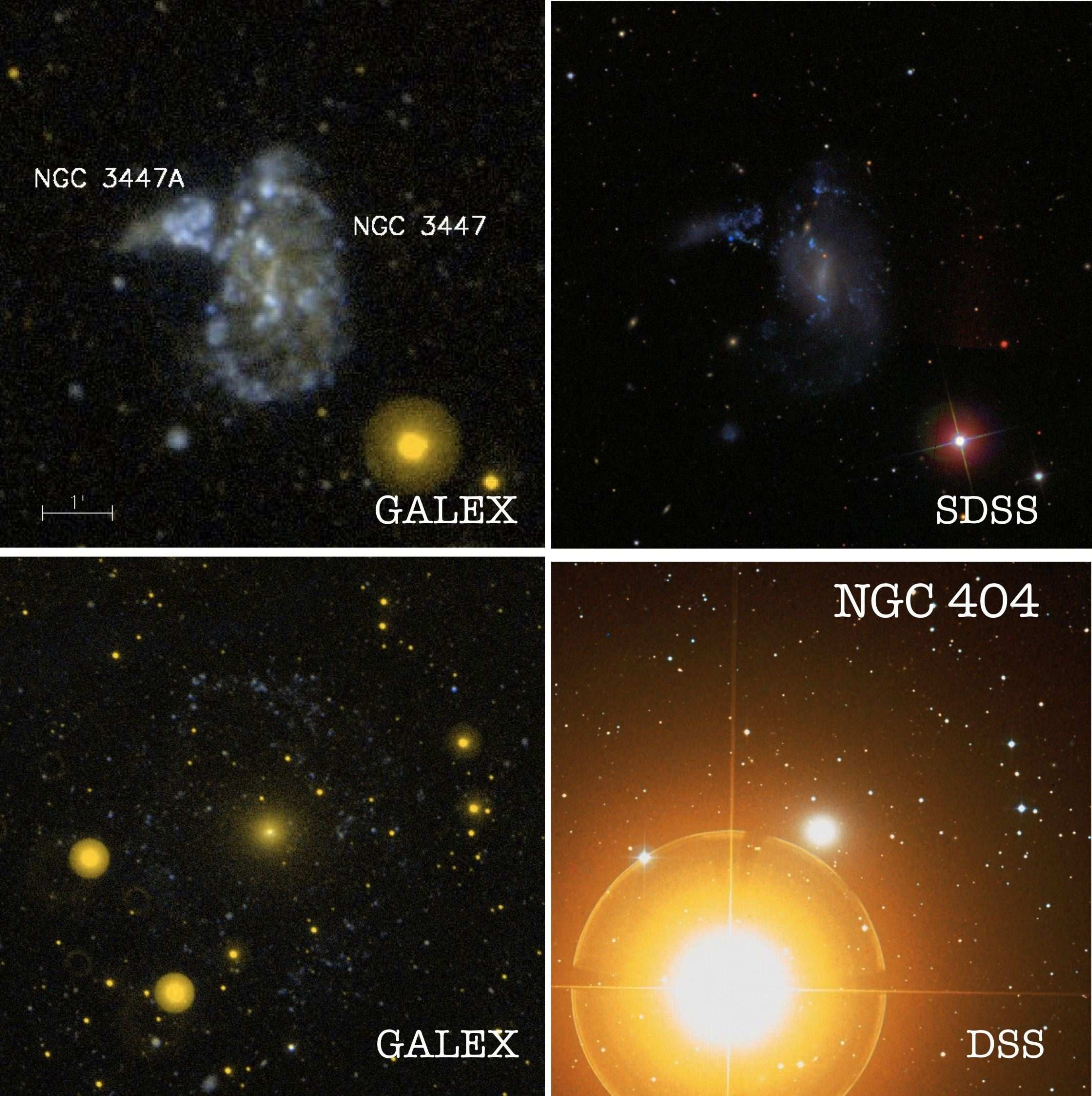}}
\caption{GALEX images (FUV: blue, NUV: yellow) and optical color-composite images of galaxies
of various types. The examples demonstrate the sensitivity of UV imaging to detect recent
star formation. One of  the most extreme examples is shown in the bottom panel: the nearest 
 S0 galaxy with a ring of FUV-emitting knots between 1 and 4 D$_{25}$ radii, from which Thilker et al. (2010)
inferred a star-formation rate 
of $\sim$ 2~10$^{-5}$ \Msun yr$^{-1}$~kpc$^{-2}$.   
 A bright foreground star (cold, visible as a yellow dot (NUV source) in the GALEX image) saturates a portion of the optical image;
the galaxy nucleus is visible in the center. 
}
\end{figure*}

\begin{figure*}
\label{f_m83}
\centerline{
\includegraphics[width=12cm]{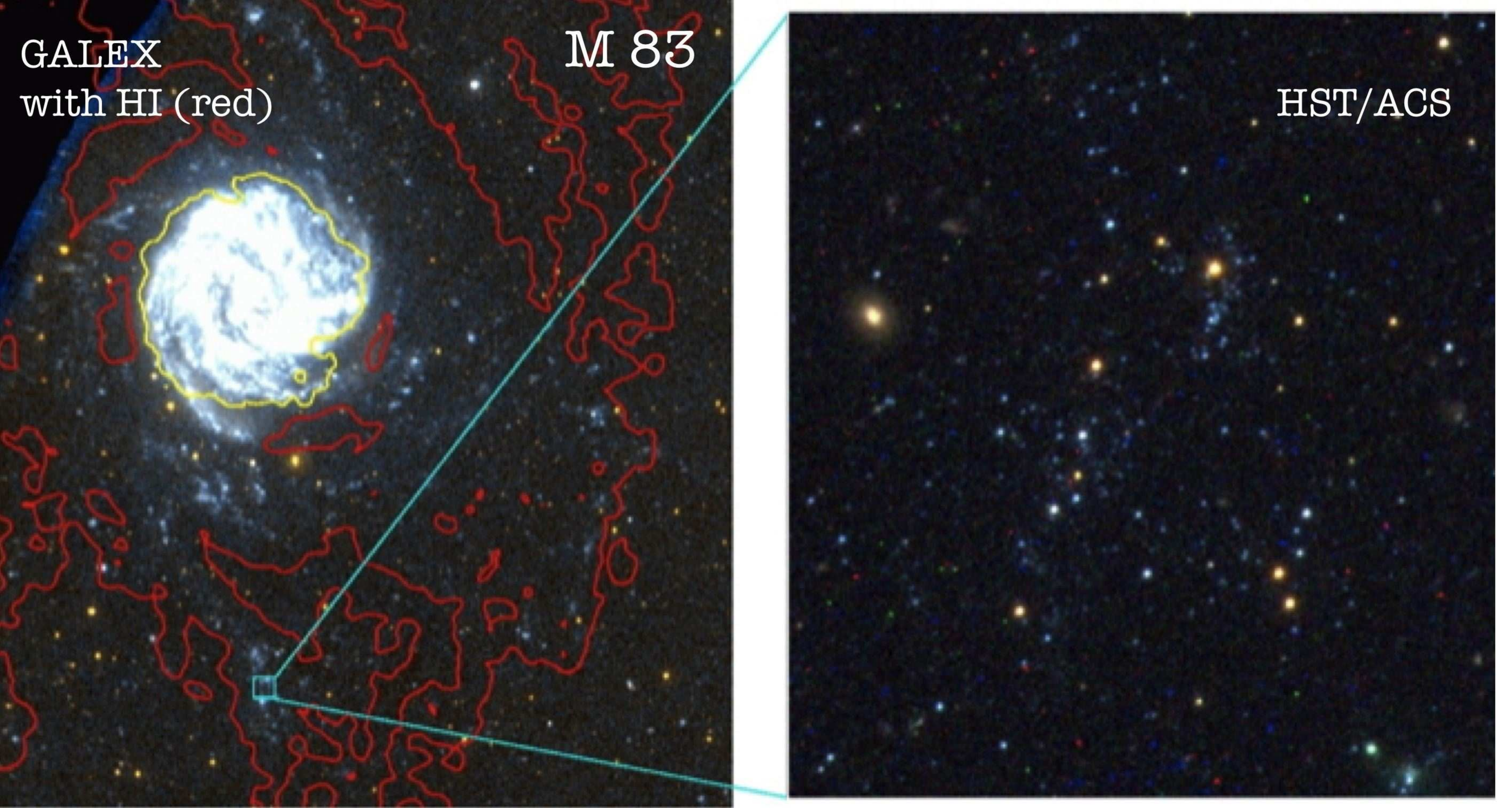}}
\caption{GALEX image (FUV: blue, NUV: yellow) 
of M83, the prototype discovery of extended UV disks (Thilker et al. 2005; another prototype was  NGC4625, Gil de Paz et al. 2005). The size of the optical image (not shown) 
is comparable to the yellow contour which is the canonical Toomre gas-density threshold for star-formation. 
HST imaging (right) resolves the UV-emitting sources in sparse hot-star associations     }
\end{figure*}

GALEX was a Small Explorer  with a data-rate higher than Hubble.
The amount of accumulated data, and 
the 
 lack of previous wide-field surveys in the UV, yielded varied and numerous results, by now too many
 to be  summarized in one paper. 
We mention here only a few highlights, some were unexpected discoveries and
 opened new lines of investigation. 

 The main science driver for  a UV sky survey, at the time when GALEX was designed (selected in 
1997), was to
measure the history of star-formation 
in the red-shift range 0-2, about 80\%
of the life of the universe, when most stars were formed, and when star-formation appeared
to have undergone significant evolution (downsizing). See e.g.  Salim et al. (2007),
 Martin et al. (2007),
Weyder et al. (2009), Marino et al. (2010, 2013),
Lee et al. (2011), Hutchings et al (2010b) to cite only a few works. Among the unexpected findings enabled by GALEX's wide-field, 
deep UV imaging, was the discovery of tenuous, extended structures of recent star formation, 
in spiral galaxies, where such extended UV disks (XUVD) stretch up to 5$\times$ the optical size of 
the galaxies (Thilker  et al 2005, 2007a; Gil de Paz 2005 and subsequent works). 
 XUVD are found in about 30\% of the galaxies. Young stellar populations, mostly in form
of FUV-bright  ring structures, were also detected in early-type galaxies, where they account in some cases
for $>$70\% of the FUV flux, though they contain only a few percent of the galaxy mass (e.g. Marino et al. 2011, 2014).

 UV fluxes are particularly sensitive to 
detect  presence of hot stars, and finely characterize them; they  trace and age-date stellar populations up to a few hundred million years old
(Bianchi 2009, 2011),
even at extremely low rates of star-formation. 
At the same time, they offer great sensitivity to interstellar dust, a key component inherently related  
to star formation. 
 In Local Group galaxies, GALEX's resolution probes 10-20~pc scales, smaller than typical
OB associations, hence uniquely tracing (and age-dating) the spatial structure of young populations throughout the whole
extent of large galaxies.
GALEX sensitive maps detect even single hot massive stars (e.g. Figure 5 of Bianchi et al. 2014b for an example in M31,
Efremova et al 2011 for NGC6822).
 A review, although quite succinct,
of 
results on star-formation from GALEX data can be found in Bianchi (2011); 
examples of the relevance of UV diagnostics 
for concurrently understanding and modeling dust and  the stellar
populations were shown by Bianchi et al. (2011c, 2014b), Kang et al. (2009). 

 The UV sky surveys also include several millions Milky Way stars. In particular, they uniquely enable
a census of hot white dwarfs, whose high temperatures and low optical luminosities make them
elusive at all wavelengths except the UV.  In low-extinction  sight-lines  (outside the Galactic plane),
GALEX MIS-depth fields can measure very hot WDs out to 20kpc (Bianchi et al. 2007).  The GALEX unbiased census 
of hot WD, an increase by over two orders of magnitude over previously known samples, opens
new ways to probe post-ABG evolutionary phases (including the poorly known initial-final mass relation,
e.g. Bianchi et al. 2011a), to find types of  binaries otherwise elusive, such as e.g. 
Sirius~B-like systems, beyond the currently known local sample (98 known $<$100pc, Holberg et al. 2013), 
to study binary evolution  (mass transfer, possible channels for SnIa progenitors) 
and Planetary Nebulae (Bianchi 2012).

\section{Looking at  the future} 

\subsection{Field-of-view versus resolution}
\label{s_galaxies}

In the past decade[s], a score of studies and results on stellar populations in nearby galaxies,
possibly more numerous than in any other area,  
have been yielded by HST and GALEX imaging (considering only the UV domain, within the scope of this 
discussion).  Such data represent two extremes, as illustrated in Fig.s 3 
and \ref{f_m83}, 
between wide-field coverage (GALEX) and very-high resolution, small field of view (HST). 
The first enables studies of stellar populations across entire extended  galaxies in the local universe, 
needed  to fully  understand the 
process and history of star formation as a function of environment (both local conditions and dynamics across the galaxy,
and external environmental factors such as interactions and gas accretion),
and to eventually build a conclusive picture. 
The high-resolution data, on the other hand, unravel the resolved stellar constituents of star-forming sites: 
and the ensemble of physical parameters derived for individual stars, in a variety of local star-forming regions 
(different metallicity, gas density, etc) provide a space- and age-tomography of such regions (e.g.,
Bianchi et al. 2014b). 
If such resolved studies coould be extended to a wide variety of conditions, with sufficient 
filter coverage to break the known [\teff, \ebv] degeneracy for hot stars, a robust calibration 
could be derived, for interpreting
 the comprehensive census of unresolved star-forming 
sites extensively mapped by the wide-field data.
In  Local Group galaxies, considering  distances  between 500 and 1000~kpc, HST imagers gives a resolution (projected) 
between  
$\sim$0.1-0.5~pc, and GALEX of $\sim$10-20~pc.    
Over 500 HST images are needed to cover one GALEX field ($\sim$700 for WFC3/UVIS, $\sim$450 for the 
largest camera, ACS/WFC, many more for ACS/HRC or STIS fields). 
The only substantial coverage of a large Local Group galaxy with HST was accomplished with the {\it PHAT} program, mapping 
one fourth of M31 (Dalcanton et al. 2012, Bianchi et al. 2014b), at the exorbitant cost of 828 HST orbits. 

The only current UV imaging capabilities with intermediate field of view and resolution 
 are SWIFT/UVOT or XMM/OM, with a limited set
of filters extending to near-UV.  Far-UV measurements are critical to characterize the hottest stars and the
youngest stellar populations.  

\subsection{Filters and sensitivity}
\label{s_filters}

The imminent launch of UVIT (Hutchings, this book) will provide a desirable bridge between HST and GALEX imaging capabilities, and 
offer a much richer choice of UV filters than GALEX. Progress in detector technology and data storage  
will enable quantum leaps  with the next generation of UV instruments. 
Field size is but one of the trade-offs. 
 The choice of filters is 
 critically  determining  to what extent the data can answer science questions. 
We may simplify a complex problem, recalling that to concurrently derive e.g. age and
extinction of stellar populations (or  \teff~ and extinction for stars), more than one color
is needed (Bianchi et al. 2014b). Metallicity is another ``free'' parameter 
(e.g., Bianchi et al. 2012). 
 In addition, the extinction curve in the UV is known to vary with environment, and the correct
solution for the physical parameters can only be achieved when such dependence will also
be conclusively constrained, and extinction properly accounted for.   A broad characterization of
extinction properties requires an adequate complement of filters. Spectroscopy provides detailed
extinction curves, but with current capabilities (HST spectrographs) it is only possible for 
a few lines of sight. 

 GALEX FUV-NUV color has the advantage to be almost reddening-free, as long as the extinction
curve is similar to the so-called ``Milky Way average'' with \rv=3.1 (Bianchi 2011). It follows that, for example,
very hot stars can be robustly selected (by \teff) from GALEX data (Bianchi et al. 2011a, 2014a).  However, in regions of 
intense star formation,  UV radiation from massive stars affects properties of local dust grains, and the
uncertainty in the extinction curve, and ultimately in the interpratation of UV fluxes,  
can be huge (example in Fig.6  of Bianchi et al. 2011c, or Table 2 of Bianchi 2011). 
 Only a sufficient complement of filters, extending to the far-UV, would allow modeling to solve concurrently for the object's 
physical parameters, and the amount {\it and type} of interstellar extinction.  
 Not a single instrument, but ideally an array of continued UV capabilities, will bring conclusive 
answers. Many are presented in these proceedings. 

 The UV-enabled characterization of young stellar populations in the local universe will inform
star-formation and galaxy-evolution models, which in turn will underpin
 interpretation of large galaxy surveys at cosmological 
distances, such as those expected from JWST  in the next decade, at high redshift, and from ground-based  facilities
at intermediate redshifts.


\section{APPENDIX. Where to find information, catalogs, and tools}

Documentation on GALEX archive data and pipeline products can be found on the MAST site\footnote{MAST database: galex.stsci.edu};
some relevant papers, science-ready catalogs and source classification can be found on 
the author's ``UVsky'' project web site\footnote{UVsky site: http://dolomiti.pha.jhu.edu/uvsky}. 

The 
BCS catalogs of GALEX unique sources are publicly available from 
the author's website\footnote{http://dolomiti.pha.jhu.edu/uvsky/BCScat}, from 
MAST\footnote{BCS on MAST: http://archive.stsci.edu/prepds/bcscat/ }
 as  part of the ``High-Level Science Product'' collection, and with SQL access from MAST casjobs 
(galex/stsci.edu/casjobs , in the context GALEXcatalogs: ``bcscat\_ais'' and ``bcscat\_mis'');
they will soon be also available in CDS's Vizier. 
Matched catalogs (GALEX BCScat's matched with SDSS, GSC2, Pan-STARRS, 2MASS, etc.), with 
useful added flags,  are being added on the http://dolomiti.pha.jhu.edu/uvsky web site,
together with science catalogs of selected samples with derived physical parameters for the sources.

The first version of the BCS catalogs (Bianchi et al. 2011a), and matched GALEX$-$optical catalogs are 
also available from http://dolomiti.pha.jhu.edu/uvsky, as well as  
from MAST\footnote{
 http://archive.stsci.edu/mast\_news.php?out=html\&desc=t\&id=378} and 
and Vizier\footnote{http://vizier.u-strasbg.fr/viz-bin/VizieR-3?-source=II/312}. 

Other GALEX catalogs  available from MAST 
include GCAT (catalog of GALEX unique
sources, not updated beyond data release GR6; it differs from BCS unique-source catalog in that 
it also includes NUV observations (up to GR6) with the FUV detector turned off, and rim/edge sources
are not eliminated, but flags are provided), and GALEX catalogs of the Kepler field.  

 Among the useful tools for exploring GALEX data at MAST, we recall ``galexview''
(galex.stsci.edu/GalexView);  since January 2014, GALEX data can also be explored within a broader context
from MAST's portal:   mast.stsci.edu/explore. 

Finally, gPhoton (Million, 
 Fleming, 
Shiao, 
2014, in preparation) is accessible at
https://github.com/cmillion/gPhoton . 

\end{document}